\documentclass[twocolumn,aps,showpacs]{revtex4}
\usepackage{graphicx,psfrag}
\def\bc{\begin{center}}
\def\ec{\end{center}}

\def\beq{\begin{equation}}
\def\eeq{\end{equation}}

\def\br{{\bf r}}

\begin{document}

\title{Two--parameter scaling theory of transport near a spectral node}

\author{Andreas Sinner and Klaus Ziegler}
\affiliation{Institut f\"ur Physik, Universit\"at Augsburg, D-86135 Augsburg, Germany}

\begin{abstract}
We investigate the finite--size scaling behavior of the conductivity in a two--dimensional Dirac electron gas
within a chiral sigma model. Based on the fact that the conductivity is a function of system size times 
scattering rate, we obtain a two--parameter scaling flow toward a finite fixed point. The latter 
is the minimal conductivity of the infinite system. Depending on boundary conditions, we also 
observe unstable fixed points with conductivities much larger than the experimentally observed values, 
which may account for results found in some numerical simulations. By including a spectral gap we extend 
our scaling approach to describe a metal--insulator transition.
\end{abstract}

\pacs{05.60.Gg, 72.10.Bg, 73.22.Pr}

\maketitle

Transport in a one--band metal is based on the dynamics of non--interacting electrons 
which are subject to random scattering. Physical quantities, such as the conductivity or 
the electronic diffusion coefficient, are obtained after averaging with respect to a random 
distribution of the scatterers. Then transport properties are controlled by large--scale
correlations in the electronic system which occur due to spontaneous symmetry breaking.
The order parameter of the latter is the average density of states \cite{wegner79}, while the
symmetry depends on the specific form of the Hamiltonian $H$. Weak fluctuations on large scales 
around the symmetry breaking saddle point are obtained by a gradient expansion, which has the action
of a nonlinear sigma model (NLSM)\cite{wegner79,wegner80,hikami80}:
\begin{equation}
 {\cal S}=\frac{1}{t}{\rm Tr}(\partial_\mu Q\partial_\mu Q)
\end{equation}
with the nonlinear field $Q$. (A symmetry--breaking term is omitted here.)
The latter is determined by the underlying symmetry of the two--particle
Green's function $G(z)G(z^*)$ with $G(z)=(H-z)^{-1}$, rather than by the symmetry of the Hamiltonian $H$
itself.

A particular class of metallic systems consists of two electronic bands with spectral nodes,
where the Hamiltonian is expanded in terms of Pauli matrices $\tau_j$.
Prominent examples are graphene \cite{geim,zhang05}, topological insulators \cite{hasan2010,zhang11} and quasiparticles
in D--wave superconductors \cite{tsvelik,ziegler} with the generic Hamiltonian 
\begin{equation}
\label{eq:Hamiltonian} 
H=H_0+V , \ \ 
H_0=h_x\tau_x+h_y\tau_y+h_z\tau_z
\ ,
\label{ham00}
\end{equation}
where $V$ is random with mean zero and variance $g$. In the special case of graphene, we have for 
the vicinity of each node $h_x=v_Fp_x$, $h_y=v_Fp_y$ with the Fermi velocity $v_F$, the components of 
the momentum $p_j$, and the gap parameter $h_z=m$. All explicit calculations will use this specific case
of $H$.
 
A number of different nonlinear fields $Q$ has been proposed for
two bands \cite{tsvelik,ostrovsky07,boquet,ziegler}. The reason for this variety of symmetry groups
is that there are actually two major approaches for studying the symmetry: Either the supersymmetry %(or replica symmetry)
is enforced by construction~\cite{efetov97} or spontaneous supersymmetry breaking is permitted~\cite{ziegler98}.

Motivated by the accurate transport measurements in graphene 
 ~\cite{geim,zhang05,kim07,elias09,bostwick09,Heer2014}, there has been much activity 
from the theoretical side to evaluate transport quantities such as the conductivity $\sigma$.
 In most calculations it is assumed that disorder is rather smooth, which implies
the absence of inter-node scattering.
Of particular interest is the size dependence, since typical graphene sample are rather small and vary in
size from sample to sample.
The behavior of $\sigma$ under a change of the linear system size $L$ has been
studied numerically \cite{bardason07,nomura07,xiong07}. There are two characteristic observations, namely 
(i) that the $\sigma(L)$ increases logarithmically with $L$ and with the disorder strength,
and (ii) that the $\beta$--function $\beta=d\log\sigma/d\log L$ is always positive but decreases 
monotonically without a finite fixed point. 
These results disagree substantially with earlier speculations on the shape of the $\beta$--function,
where two finite fixed points were proposed \cite{ostrovsky07}. Given the fact that
there is a very robust minimal conductivity $\sigma^{}_{\rm min}\propto e^2/h$ in the experiments, it is
rather surprising that the numerical calculations do not indicate the existence of a finite fixed point 
for the conductivity. This might be a hint that the simulations have not reached the asymptotic regime.

In the following we assume weak and slowly varying disorder so that there is no scattering between different spectral nodes.
Then we briefly discuss the realization of the chiral sigma model (CSM) with broken supersymmetry for 
a two-band system of Ref.~\cite{ziegler} and evaluate the corresponding finite--size scaling of the conductivity. 
Although the $\beta$--function is sensitive to the existence or absence of a zero mode in the finite 
system, it always describes a flow towards a finite attractive fixed point that agrees with the minimal 
conductivity at the Dirac node. This provides a surprisingly simple two--parameter scaling picture for 
transport in two--band metals with a spectral node. 

There are several options to evaluate the transport properties at the Dirac node. One is based on the diffusion coefficient
\begin{equation}
%D_0(E)=\lim_{\epsilon\to0}\epsilon^2{\rm Tr}\sum_\br  r_k^2\langle G_{\br0}(E+i\epsilon)G_{0\br}(E-i\epsilon)\rangle_d
D_0=\lim_{\epsilon\to0}\epsilon^2\sum_\br  r_k^2{\rm Tr}_2\langle G_{\br0}(i\epsilon)G_{0\br}(-i\epsilon)\rangle_d
\ ,
\label{diff0}
\end{equation}
another one is provided by the Kubo formula of the conductivity as
\begin{equation}
\sigma(\omega)=-\frac{e^2\omega^2}{2h}\sum_\br  r_k^2
%\langle G_{{\bar r}{\bar r}'}(\omega/2+i\epsilon)G_{{\bar r}'{\bar r}}(-\omega/2-i\epsilon)\rangle_d
{\rm Tr}_2\langle G_{\br0}(\omega/2)G_{0\br}(-\omega/2)\rangle_d
\label{kubo0}
\end{equation}
for the response to an external electromagnetic field with frequency $\omega$. 
${\rm Tr}_2$ is the trace with respect to the Pauli matrices of the two--band Hamiltonian.
These expressions are connected by the analytic continuation $\epsilon\to i\omega/2$.
The correlation function
\begin{equation}
K_{\br\br'}=\langle G_{\br\br'}(i\epsilon)G_{\br'\br}(-i\epsilon)\rangle_d
\ ,
\label{corr0}
\end{equation}
which appears in both expressions from the average $\langle ...\rangle_d$ with respect to random scatterers, 
is available from a field--theoretical calculation~\cite{ziegler09}.
This is based on the symmetry relation $-\tau_xH_0^*\tau_x=H_0$ of the two--band Hamiltonian.
A consequence is that the energy eigenfunction in the upper and the lower band are related as
$\Psi_{-E}=\tau_x \Psi_E^*$. It allows us to write
\beq
G(-i\epsilon)=(H_0+V-i\epsilon)^{-1}=-\tau_x(H_0^*-V+i\epsilon)^{-1}\tau_x
\ ,
\eeq
% where $\tau^{}_x$ is areal non--diagonal Pauli matrix, 
which implies the chiral symmetry $e^{\hat S}{\hat H}e^{\hat S}={\hat H}$ for
\beq
{\hat H}=\pmatrix{
H_0+V  & 0 \cr
0 & H_0^* -V \cr
} , \ \ \ 
{\hat S}=\pmatrix{
0 & \varphi\tau_x \cr
\varphi'\tau_x & 0 \cr
} 
\ .
\eeq
The symmetry group depends only on the single pair of Grassmann variables $(\varphi,\varphi')$.
Thus, the nonlinear field is $Q=e^{\hat S}$ and we can write
\begin{equation}
K_{\br\br'}
=\frac{4\eta^2}{g^2}
\frac{1}{\cal N}\int \varphi_{\br}\varphi_{\br'}' e^{-{\cal S}_2}{\cal D}[\varphi] , \ 
%S_2=\frac{4\eta}{g}\int_q(\epsilon+D q^2)\varphi_{q}\varphi_{-q}'
{\cal N}=\int e^{-{\cal S}_2}{\cal D}[\varphi]
\label{corr1}
\end{equation}
with the bilinear CSM action
\begin{equation}
{\cal S}_2=\frac{4\eta}{g}\left[\epsilon {\rm Tr}(\varphi\varphi')+D{\rm Tr}(\partial_\mu\varphi\partial_\mu\varphi')\right]
%(\varphi_{+,q}\varphi_{+,-q}'+\varphi_{-,q}\varphi_{-,-q}')
\ .
\label{action2}
\end{equation}
It should be noticed that the bilinear form is characteristic for the Dirac node.
There are also quartic terms away from the node \cite{PRB86}. $D$ is the (renormalized) diffusion coefficient
\begin{equation}
D=\frac{\eta g}{2}{\rm Tr}\sum_\br  r_k^2 {\bar G}_{\br0}(i\eta){\bar G}_{0\br}(-i\eta) 
\label{diff2}
\end{equation}
with the effective Green's function ${\bar G}(z)=(\langle H\rangle_d-z)^{-1}$.
The definition of the diffusion coefficient $D_0$ in Eq. (\ref{diff0}) and the correlation function in Eqs. 
(\ref{corr0}), (\ref{corr1}) imply the relation $D_0=2\eta D/g$. Moreover, by comparing the NLSM action
of a one--band metal with the CSM action ${\cal S}_2$ we get for their prefactors the relation 
\begin{equation}
t^{-1}\longleftrightarrow 4\eta D/g=2 D_0,
\label{diff3}
\end{equation}
which is the conductivity due to the Einstein relation
$\sigma=2 e^2 \eta D/gh=e^2 D_0/h$. This can be used now to calculate the $\beta$--function
from $D_0$, in analogy with the treatment of a one--band metal. 
In order to determine the size dependence of $D_0$ we use a simple approximation for a first estimate and
in a second step a more detailed numerical summation of $D$ in Eq. (\ref{diff2}).

\begin{figure*}[t]
\includegraphics[width=7cm]{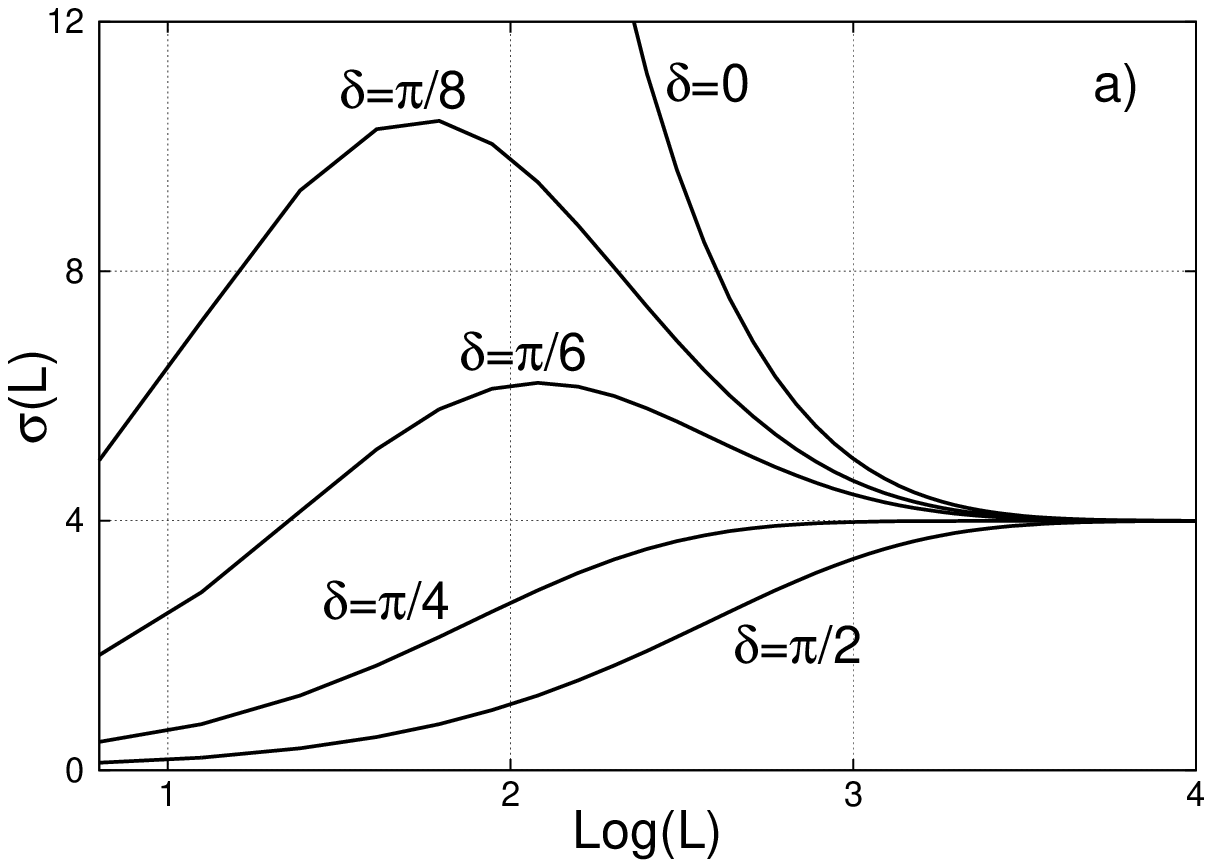}
\hspace{4mm}
\includegraphics[width=7cm]{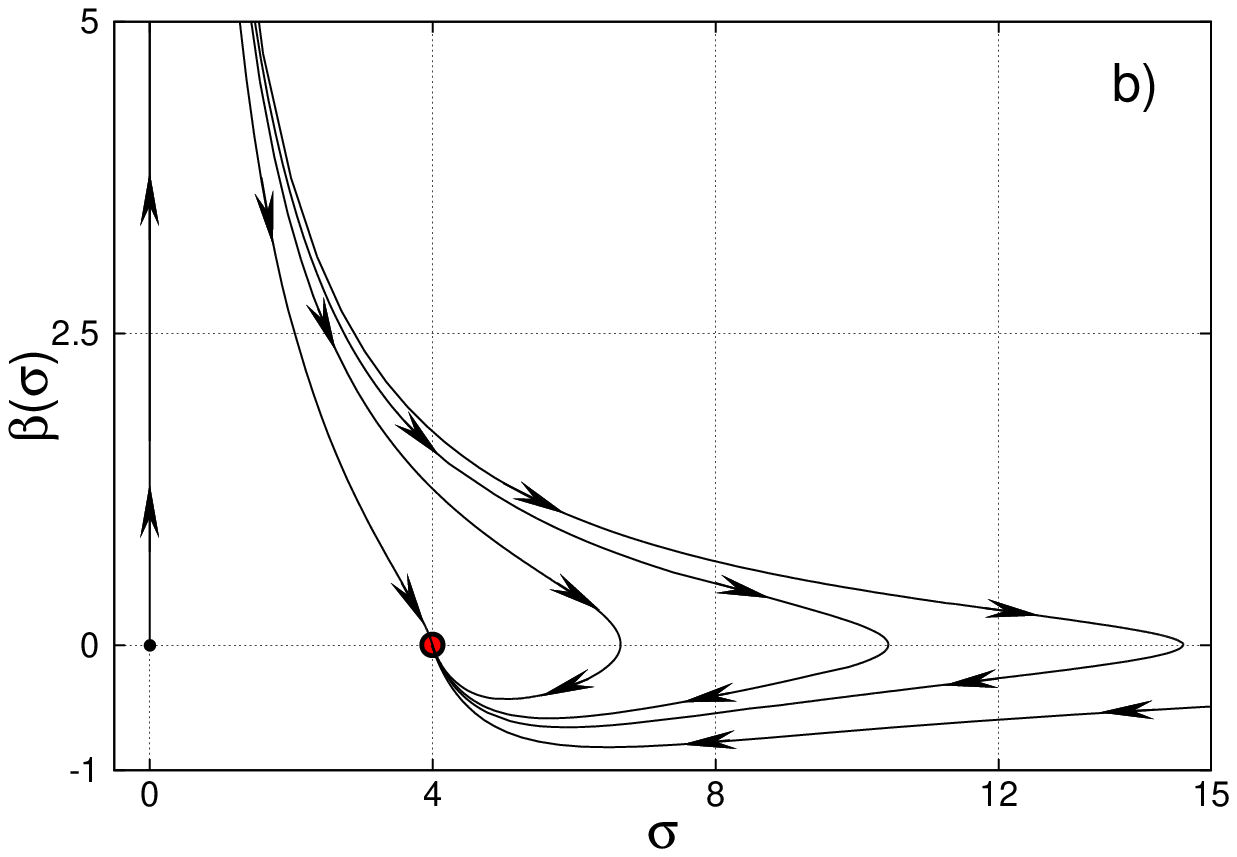}
\caption{a) Conductivity in units of $e^2/\pi h$ as a function of length $L$ at fixed 
scattering length $\eta\sim0.2$, calculated for different BC (i.e., different values of $\delta$). 
b) %Schematic %flow of the 
The $\beta$--function 
of the weakly disordered massless Dirac electron gas in two dimensions for 
different BC. There is a universal attractive fixed point at $\sigma= e^2/\pi h$.}
\label{fig:cond}
\end{figure*}

For a finite sample of size $L\times L$ and no gap ($m~=~0$) the main effect on $D$ is an infrared cut--off in the
Fourier integral, assuming that the largest wavelength is $L$:
\beq
\sigma\approx\frac{2\eta^2}{\pi}\int_{L^{-1}}^\infty \frac{k~dk}{[k^2+\eta^2]^2}
\sim \frac{1}{\pi}\left(1-\frac{1}{\eta^2L^2}\right)
\ .
\label{cond2a}
\eeq
%for $\lambda^{-1}\eta\gg \eta L\gg 1$. 
for $\eta L\gg 1$.
%The upper cut--off $\lambda$ corresponds with the bandwidth. 
This result indicates 
that the conductivity increases monotonically with the size and its $L$--dependence scales with the 
scattering rate $\eta$: $\sigma(L,\eta)=\sigma(\eta L)$. Moreover, the $\beta$--function reads 
in this approximation
\beq
\beta\sim 2\pi(1-\pi\sigma)
\ ,
\label{beta1}
\eeq
which has a fixed point $\sigma^*=1/\pi$ in units of $e^2/h$. This is the well--known minimal
conductivity $\sigma_{\rm min}$ of Dirac fermions.
Although this approximation is reliable near the fixed point, it may not be so good further
away from the fixed point. The reason is that we have not considered (i) that the spectrum of
a finite system is discrete and (ii) that the boundary conditions can be crucial. The effect
of the latter is know to be important, for instance, in graphene ribbons, because the system may or may
not have a gap \cite{BreyFertig2007,Akhmerov2007,Enoki2010}.

The discrete spectrum of the gapless Dirac Hamiltonian $H_0$ in Eq. (\ref{ham00}) is $E=\pm\sqrt{k^2_n+k^2_m}$
 with wave numbers $k^{}_j=2(\pi j+\delta)/L$, $j=0,\pm1,\pm2,\pm3,...$.
%,\pm L/2$. 
The parameter $\delta$ depends on the boundary condition (BC). In particular, we have $\delta=0$ for periodic BC and
$\delta\neq 0$ for BC with a phase shift $\delta$ of the wave function at opposite boundaries. Thus, only
$\delta=0$ has a zero mode, whereas $\delta\ne0$ has a spectral gap that increases with increasing $\delta$. 
This mimics the situation of the tight--binding model in the case of a graphene ribbon, where armchair (zigzag)
boundaries provide a gapless (gapped) spectrum~\cite{BreyFertig2007,Akhmerov2007,Enoki2010}. 
With this discrete spectrum we calculate the conductivity in Eqs.~(\ref{diff2}) and (\ref{diff3}) as a function 
of size $L$ with generic BC, characterized by the phase shift $\delta$, at the Dirac point ($E=0$):
\begin{equation}
\label{eq:CondDescr}
\sigma(\eta,L) = \frac{4\eta^2}{L^2}\sum_{n,m}~\frac{1}{[k^2_n+k^2_m+\eta^2]^2}
\ .
\end{equation}
The sum converges and gives us a conductivity that depends only on $\eta L$.
$\sigma(\eta L)$ is plotted in Fig.~\ref{fig:cond}a, where for $\eta L\sim \infty$ its value
agrees with the minimal conductivity of Eq.~(\ref{cond2a}).
For intermediate values $\eta L$, on the other hand, the conductivity depends
strongly on the parameter $\delta$, though. In the case of periodic BC ($\delta=0$) the behavior is dominated by 
the zero energy mode. Its contribution to the conductivity decreases as $\sim L^{-2}$ with increasing
sample size, and the conductivity represents a monotonically decreasing function of the length $L$. 
For $\delta\geq \pi/4$ the zero mode is strongly suppressed. In this case the conductivity increases 
monotonically with $\eta L$ (cf. Fig. \ref{fig:cond}a). In particular, there is a relatively 
broad regime where it grows logarithmically with 
$\eta L$, i.e.$\sim{const}+2.5\ln(\eta L)$, which agrees with known analytical~\cite{boquet,Senthil1999} 
and numerical~\cite{bardason07,nomura07} results. 
Finally, there is an intermediate regime for $0< \delta < \pi/4$, in which the
conductivity increases up to a maximum and then approaches the asymptotic minimal conductivity from above. 
 
%{\it Calculation of the $\beta$--function:}-- 

\begin{figure}[t]
\includegraphics[width=7.5cm]{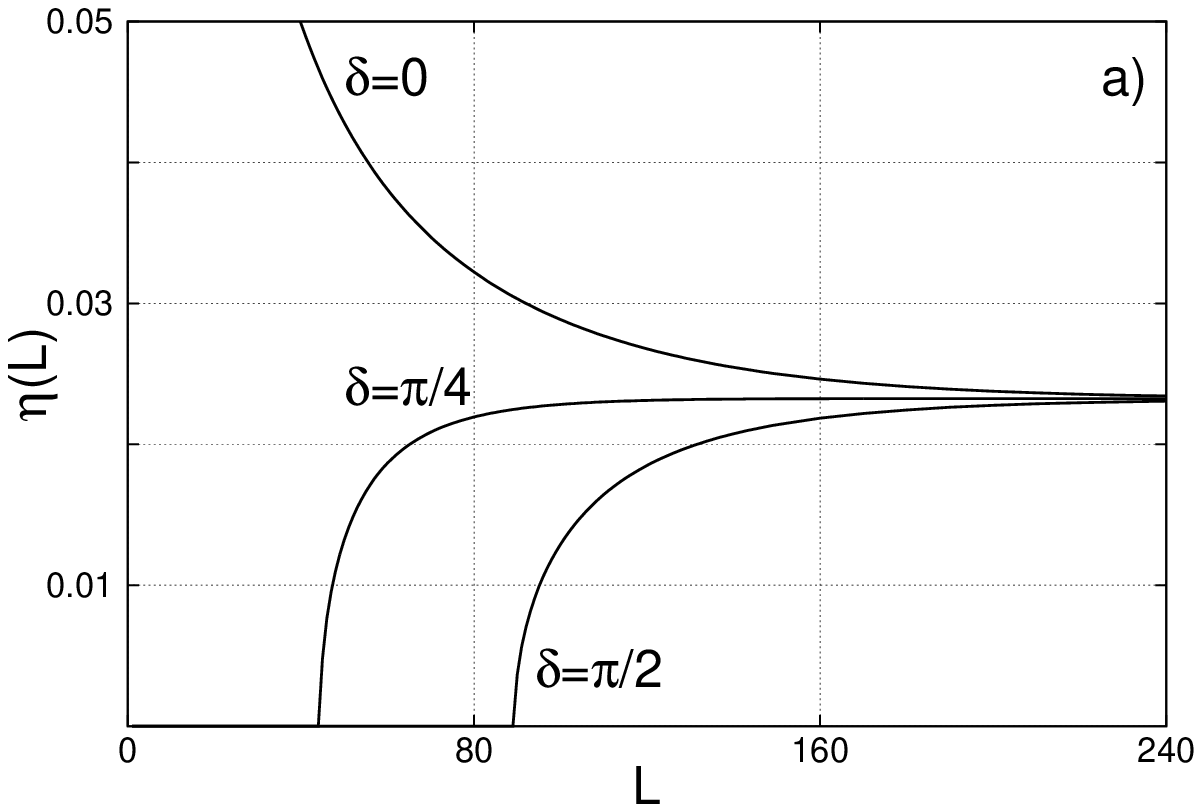}
%\hspace{5 mm}
\includegraphics[width=7.5cm]{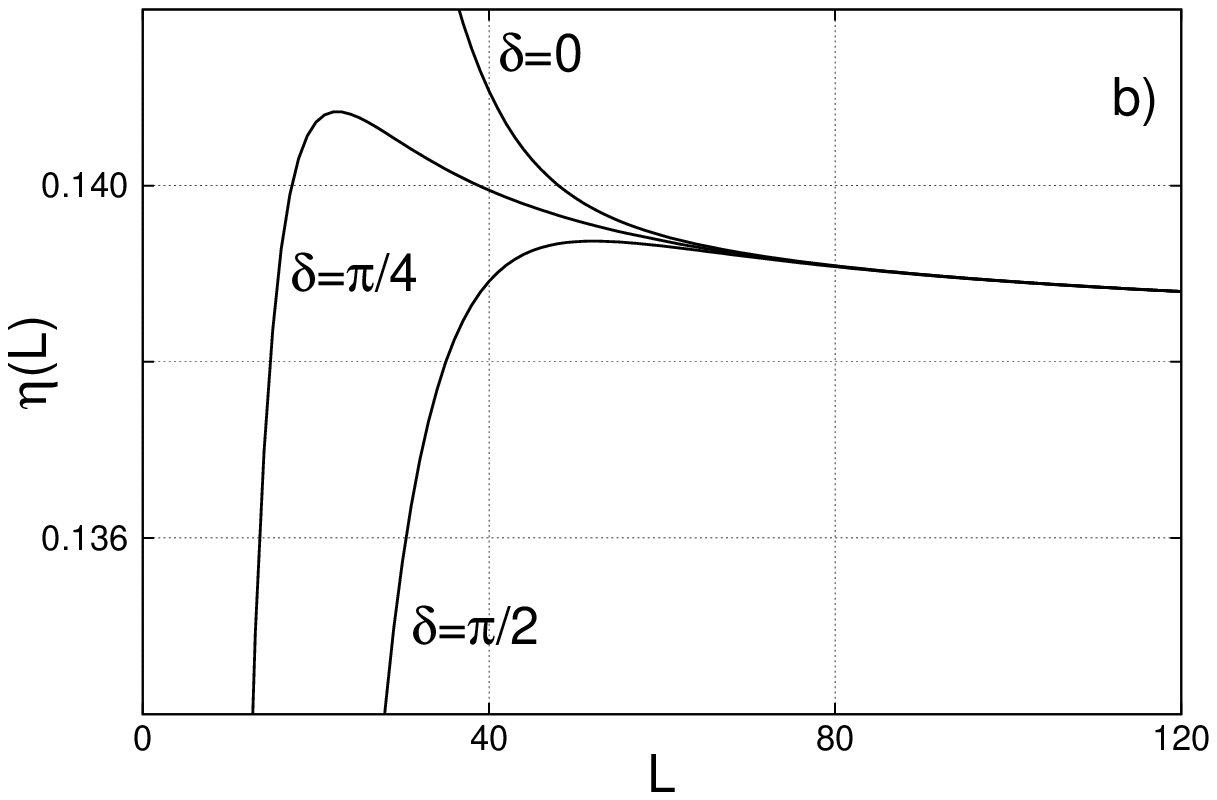}
\caption{
The behavior of the scatterig rate calculated from Eq.~(\ref{eq:SaddlePoint}) for two fixed values of the disorder strength $g=1.1$ a) and $g=1.6$ b), for  three different values of $\delta$ each. 
}
\label{fig:ScatRate}
\end{figure}

The scattering rate $\eta$, which so far appeared in the conductivity as a free parameter, can also be calculated
as a function of system size $L$ and disorder strength, using the self--consistent Born approximation~\cite{ando1998,Fradkin1986}
\begin{equation}
\label{eq:SaddlePoint}
\frac{1}{g} = \frac{1}{L^2}\sum^{+L}_{n,m=-L}~\frac{1}{k^2_n+ k^2_m +\eta^2}.
\end{equation} 
The calculation for a finite sample is again a sum over the discrete wave numbers $k_j$, in analogy to
the calculation the conductivity, and gives a non--monotonous scattering rate with respect to $L$
that increases up to a certain length and approaches asymptotically a finite value. The asymptotic value depends
on $g$ but is indifferent to $\delta$. The way $\eta$ approaches this value depends significantly on $\delta$, though: It decreases with increasing $g$ and decreasing $\delta$, cf.~Fig.~\ref{fig:ScatRate}.

\begin{figure}[t]
\includegraphics[width=7.8cm]{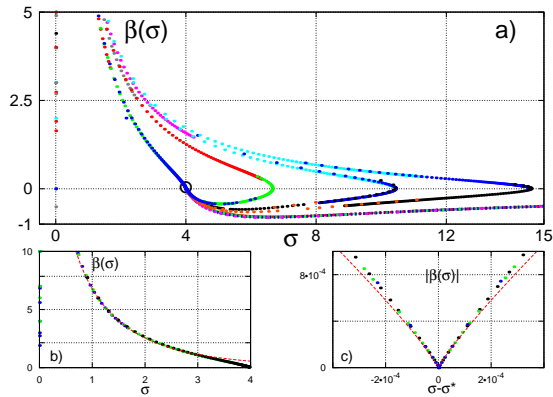}
\caption{(Color online) 
a) The same $\beta$--function as in Fig.~{\ref{fig:cond}a} represented as the scaling plot for different 
strengths of disorder. Different branches are calculated with (clockwise) $\delta=\pi/2$ (antiperiodic BC), 
$\delta=0.16\pi$, $\delta=\pi/8$, $\delta = 0.101\pi$ and $\delta=0$ (periodic BC). The disorder strength 
varies within $g\in[0.5,2]$; pieces calculated for different disorder are depicted in different colors. 
b) A particular branch of the $\beta$--function calculated with $\delta=\pi/4$ and fitted with 
the conductivity formula $\sigma(L)=k \log[1+u^2\log^2L]$ from Ref.~\cite{PRB86} with $k=1$ and $u=7.5$ 
(dashed line). c) The absolute value of the $\beta$--function in the vicinity of the fixed 
point calculated with $\delta=\pi/2$ (left branch) and $\delta=0$ (right branch). The red dashed line represents 
a fit with the formula $|\sigma-\sigma^{\ast}_{}|^{0.875}$ (shown here only as a guide to the eye) which 
is the best match in this parametric area. 
%The number in the exponent is not universal but approaches slowly %the unity for 
%even smaller $|\sigma-\sigma^{\ast}_{}|$.
}
\label{fig:betaf1}
\end{figure}

Once the $L$--dependence of the scattering rate is taken into account the $\beta$--function for 
different values of $\delta$ and $g$ is calculated from Eq.~(\ref{eq:CondDescr}).
Plotting the curves for different values of $g$ together, the graphs collapse on a single curve, as 
depicted in Fig.~{\ref{fig:betaf1}a}. Moreover, regardless of the parameters, all solutions are attracted to the 
fixed point $\sigma^\ast$ with the value of the minimal conductivity. However, there are two types of 
$\beta$--functions, one that approaches the fixed point from positive values (like the approximation
in Eq.~(\ref{beta1})) and another one from negative values. The positive branch of the $\beta$--function
coincide with $\delta>\pi/4$, while the negative branch is associated with $\delta<\pi/4$.
The negative branch also starts for small $L$ with positive values of the $\beta$--function and reaches an
unstable fixed points at values much larger than the experimentally 
observed conductivity. However, the $\beta$--function does not stop there but keeps flowing toward 
the only attractive fixed point at the observable value $\sigma^{\ast}_{}$.
Thus, the BC related parameter $\delta$ enforces the two--parameter scaling, whereas for fixed $\delta$ we
obtain the one--parameter scaling. In particular, the positive branches resemble the numerically evaluated 
$\beta$--functions found in~\cite{bardason07} and~\cite{nomura07}. 
Moreover, the main part of these branches is fitted excellently with the double logarithm formula obtained 
in leading order of perturbation theory in Ref.~\cite{PRB86}, cf. Fig.~{\ref{fig:betaf1}b}. 
However, the limitation of the one--loop approximation does not allow to approach analytically the quasi--fixed point 
at which the $\beta$--function changes the sign.

Close to the fixed point, the $\beta$--function exhibits a 
%general, i.e. boundary conditions independent, 
power law behavior $\beta(\sigma)\sim|\sigma-\sigma^{\ast}_{}|^y$, with an exponent $y$ that
approaches unity for very small deviations from the fixed point, in agreement with the approximation
in Eq.~(\ref{beta1}). For $|\sigma-\sigma^{\ast}_{}|/\sigma^{\ast}_{}\sim 10^{-4}$ we can fit our
curves with $y\sim7/8$, as it is shown in Fig.~{\ref{fig:betaf1}c}.
This is a crossover to asymptotic power law with exponent 1, which might be important for comparison
with numerical simulations and experimental measurement. For the latter we have typical values of
%the scattering rate 
$\eta \approx 0.7 ...70$ meV \cite{kim07,pallecchi11} and typical sizes $L\approx 10^{-4}$~m
\cite{castroneto09,andrei11} such that we get $\eta L/v_F\hbar\approx 7\cdot 10^2 ..7\cdot 10^5$, which matches well the parametric regime 
$|\sigma-\sigma^{\ast}_{}|/\sigma^{\ast}_{}\sim10^{-4}$.

\begin{figure*}[t]
\includegraphics[height=5.5cm]{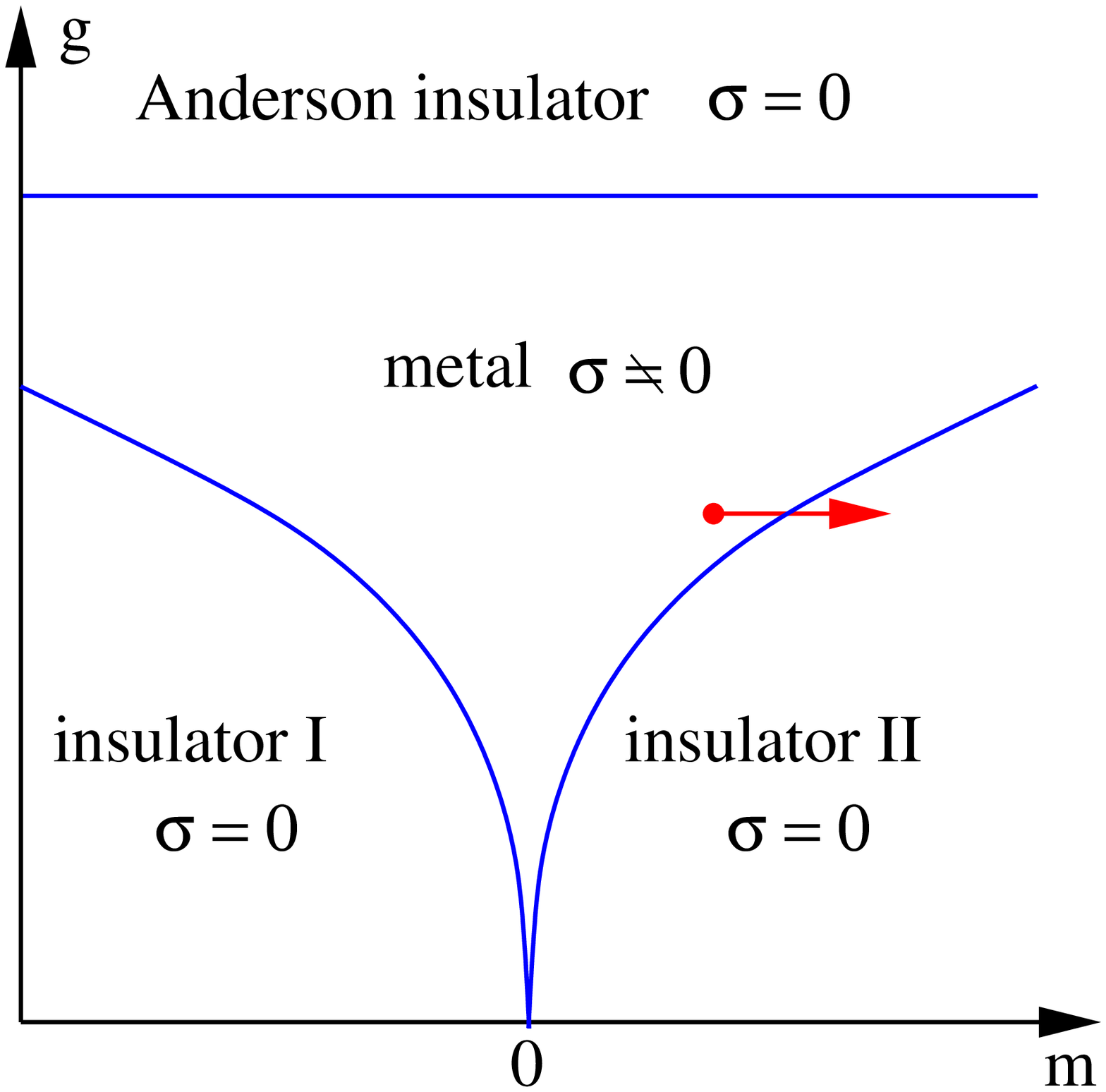}
\hspace{10 mm}
\includegraphics[height=5.1cm]{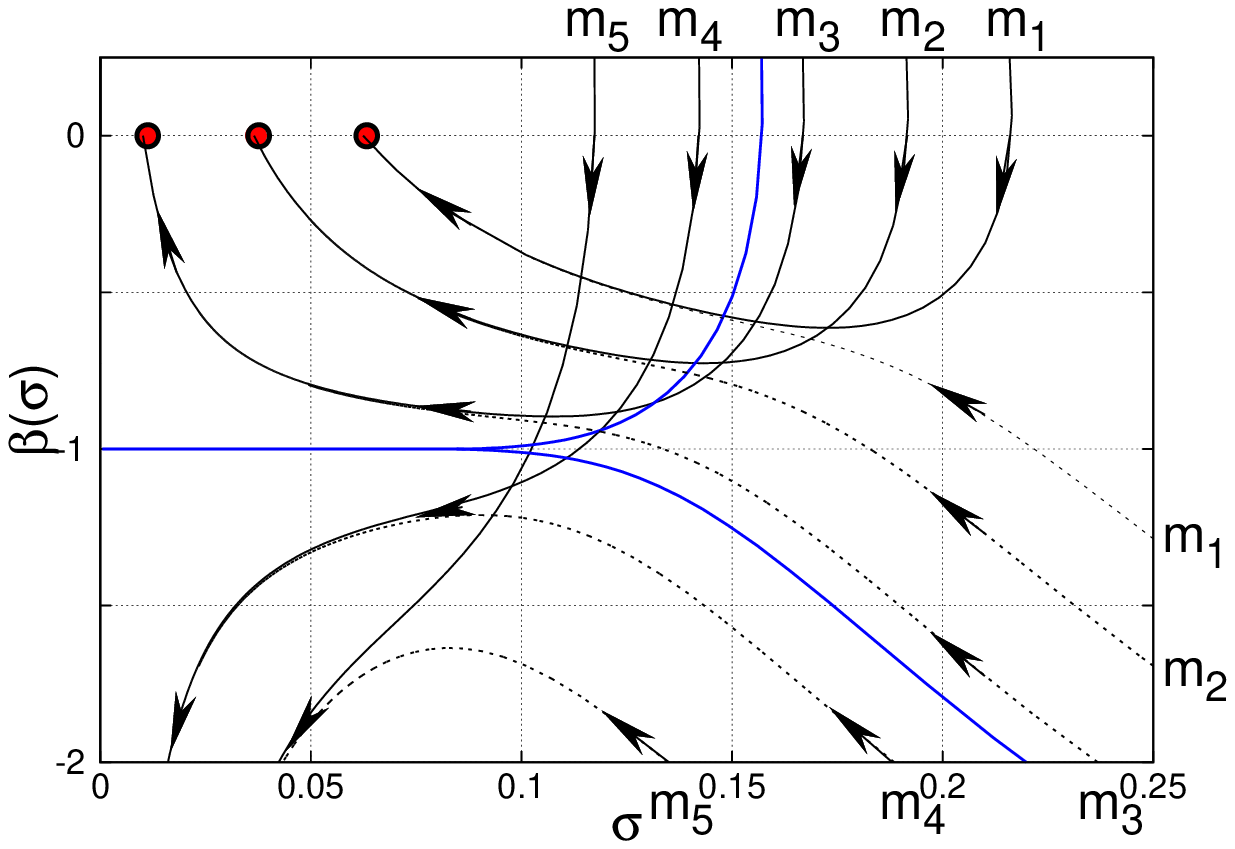}
\caption{(Color online) Metal--insulator transition in the gapped disordered 2D Dirac electron gas.
Left: Qualitative phase diagram of the metal--insulator transition. Red arrow marks the line of constant disorder strength 
along which we cross the separation line between metallic and isolating phases. 
Right: Scaling flow of the conductivity along the red arrow. Solid lines represent the $\beta$--function  
calculated with $\delta=\pi/2$, the dashed lines those with $\delta=0$. 
The disorder strength is fixed at $g=2$ and the gap 
is equidistantly changed from $m^{}_1=0.301$ to $m^{}_5=0.305$. There 
is a critical gap $m^{}_c\sim 0.30339$ (blue line), separating
the conducting from the insulating regime. The fixed point of the conducting regime is shifted with $m$ 
toward zero, thus suggesting the second order phase transition scenario. 
}
\label{fig:mass3d}
\end{figure*}

{\it Metal--insulator transition in the gapped disordered 2D Dirac electron gas:}--
Returning to the Hamiltonian in Eq.~(\ref{ham00}) we include now a gap term $m\tau_z$.
This would allow us to study a metal-insulator transition (MIT), as predicted earlier
in the literature \cite{ziegler09,Medvedeva2010}. The possibility of tuning the gap 
experimentally in a sample with a particular disorder configuration, for instance, by controllable 
hydrogenation \cite{elias09,bostwick09}, provides a transition at fixed disorder strength 
by varying the gap: For $0<m<m_c$ we have a metal and for $m\ge m_c$ a band insulator.  
On the level of practical implementation, the gap is built into 
Eq.~(\ref{eq:CondDescr}) and in the self--consistent Born approximation
%(\ref{eq:SaddlePoint}) 
by replacing $k^2_n+k^2_m$ with $k^2_n+k^2_m+m^2$. Then both, the scattering rate and the conductivity,
depend also on $m$
%yet another dimensionless parameter $mL$, 
which describes the $\beta$--function $\beta(\sigma,\delta,m)$ in a 3D parametric space, as 
illustrated in Fig.~\ref{fig:mass3d}. The fixed point $\sigma^*$ turns out be unstable with
respect to the variable $m$: Gradually increasing the gap from zero upward we observe 
a shift of the fixed point from $\sigma^*=1/\pi$ toward zero. At zero a 
critical gap is reached, where the critical value $m_c$ depends on the disorder strength. 
For $m>m_c$ the system does not have any fixed points with finite conductivity but undergoes a transition
to the insulating phase. For a broad range of disorder strengths it is verified that the 
asymptotical behavior of the $\beta$--function on the critical trajectory is $\beta\sim1$.

{\it Conclusions:}-- A number of experimental 
investigations~\cite{geim,zhang05,kim07,elias09,bostwick09,Heer2014} provides strong 
evidence for a universal, sample shape and disorder strength independent conductivity of a weakly 
disordered 2D Dirac electron gas. This apparently contradicts to claims of some numerical~\cite{bardason07,nomura07} and analytical~\cite{boquet,ostrovsky07,Senthil1999} work, 
which predict a supermetallic fixed point at infinite conductivity.
% with logarithmic conductivity growth. 
In this work we have investigated the conductivity within the CSM approach~\cite{ziegler09} and found
that the conductivity can indeed flow to (unstable) fixed points at values much larger than 
the experimentally observed conductivity. However, the $\beta$--function does not stop there but 
keeps flowing back to smaller conductivities to reach eventually the attractive bulk fixed point 
at $\sigma^{\ast}_{}=1/\pi$ in units of $e^2/h$. The details of this flow depend crucially on the boundary condition. The conductivity $\sigma$ depends on the scattering rate $\eta$ and the system length
$L$ as $\sigma(\eta,L)={\bar\sigma}(\eta L)$.
A spectral gap shifts the fixed point $\sigma^*$ to smaller values, indicating an unstable fixed
point against gap opening. This leads eventually to a metal--insulator transition.

\end{document}